\documentclass[12pt]{revtex4}
\usepackage{graphics,epsfig,graphicx}
\usepackage{color}
\usepackage{makeidx}
\usepackage{latexsym}

\begin{document}

\title{ 1D exciton band and exciton-phonon interaction in a single polymer chain}

\author{ F. Dubin$^*$, J. Berr\'{e}har, R. Grousson, T. Guillet,  C. Lapersonne-Meyer,\\ M. Schott, V. Voliotis \\
{\em Groupe de Physique des Solides, UMR 7588 of CNRS,} \\
{\em Universit\'{e}s Pierre et Marie Curie et Denis Diderot, Tour 23, 2 place Jussieu,} \\
{\em 75251 Paris cedex 05, France }}


\begin{abstract}
The excitonic luminescence of an isolated polydiacetylene polymer
chain in its monomer matrix is studied by micro-photoluminescence.
These chains behave as perfect 1D excitonic systems with the
expected 1/$\sqrt{E}$ density of states between 5 and 50 K. The
temperature dependence of the homogeneous width is quantitatively
explained by interaction with longitudinal acoustic phonons of the
crystal in the range of temperature explored.
\end{abstract}

\maketitle

\begin{figure}
\begin{center}
\mbox{\includegraphics[width=5 cm,height=9 cm,angle=-90]{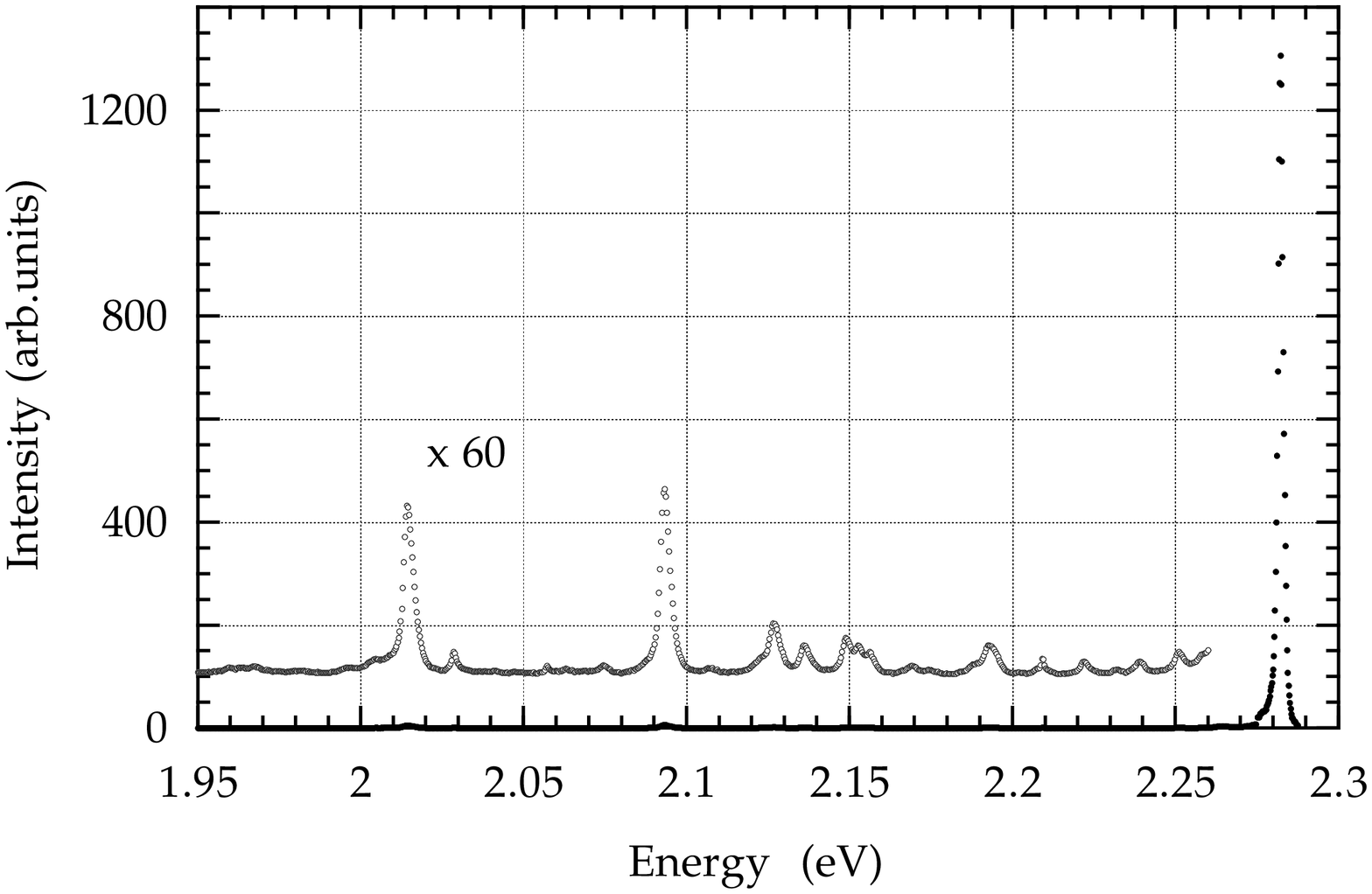}}
\caption{Fluorescence emission spectrum of an ensemble of isolated
red chains excited at 502 nm at 15 K. Full circles: zero phonon
line at 2.28 eV. Open circles: vibronic emission spectrum; the two
most intense vibronic lines correspond to the C=C stretch (D line
at 2.09 eV) and C$\equiv$C stretch (T line at 2.015 eV) (From
\cite{Lecuiller2002})} \label{lumspectrum}

\end{center}

\end{figure}

Conjugated polymers contain delocalized $\pi$ electrons from
conjugated unsaturated C-C bonds. They behave as organic
semiconductors with large gap. They are usually highly disordered
samples \cite{Schott}, but making use of the special
polymerization mechanism of polydiacetylenes (PDAs)
\cite{formule3B} a model system has been obtained in order to
study the electronic properties of conjugated polymers.
Diacetylenes are polymerized in the solid crystalline phase and by
a judicious choice of side groups \cite{formule3B} and
polymerization conditions, and one can obtain a monomer single
crystal containing a very low concentration of chains of the
corresponding PDA. These chains  are highly ordered, non
interacting, linear and very long \cite{Spagnoli}. These PDAs are
organic large gap semiconductors where the principal electronic
excitation is an exciton having most of the oscillator strength
and a large binding energy of 0.5 eV \cite{Horvath}. Bulk PDAs are
known to exist in two electronic structures called "red" and
"blue" phases \cite{Eckhardt,Koshihara} with an intense excitonic
absorption around 2.4 and 2 eV respectively. In 3BCMU
\cite{formule3B} monomer crystals both types of chains coexist and
exhibit an excitonic resonance fluorescence \cite{Lecuiller98}.
Blue chain fluorescence is very weak whereas red chains have a
high fluorescence quantum yield of 0.3 at 15 K \cite{Lecuiller99}.
The present work is exclusively concerned with red chains. The
luminescence spectrum of red chains exhibits an intense zero
phonon line and several much weaker vibronic replicas (see Figure
\ref{lumspectrum}). The zero phonon line is centered at 2.28 eV at
low temperature. The two main vibronic peaks correspond to the
stretching of the C$=$C and C$\equiv$C bounds and will now be
denoted by D and T respectively. These two lines are centered at
2.09 and 2.01 eV and are due to radiative recombination with
emission of a chain optical phonon of the appropriate momentum
\cite{Lecuiller98}.

The very high dilution of red chains and their high fluorescence
yield allow the study of a \underline{single} red chain by
micro-photoluminescence ($\mu$-PL) experiments \cite{Guillet2001}.
The zero phonon line-shape is Lorentzian and much broader than
calculated from the exciton lifetime \cite{Lecuiller2002}. The
vibronic emission lines are broader and asymmetric. These
line-shapes are analyzed below. It is shown that a purely 1D
exciton density of states (DOS), i.e with its 1/$\sqrt{E}$
singularity, quantitatively accounts for the vibronic line-shapes.
This analysis requires the excitons to be in thermal equilibrium
with the surrounding 3D monomer crystal lattice. Indeed, the
temperature dependence of the zero phonon line-width is
quantitatively explained by a 1D exciton-3D LA phonon interaction
strong enough to ensure that thermal equilibrium.

The 3BCMU crystals analyzed were identical to the ones described
in \cite{Guillet2001}, i.e with a concentration of red chains
smaller than $10^{-8}$ in weight. The excitation wavelength of the
$Ar^{+}$ laser was chosen at 497 nm, nearly resonant to one
vibronic absorption line. The excitation power was below 1 $\mu$W
to keep the measurement in the low excitation regime, i.e with at
most one exciton per chain. The excitation laser beam was focused
on the sample using a microscope objective with a numerical
aperture of 0.6 yielding a diffraction limited laser spot of
$\approx$ 1 $\mu$m diameter. The signal was analyzed through an
imaging spectrometer coupled to a $N_{2}$ cooled CCD camera. The
spectral resolution was about 100 $\mu$eV.

\begin{figure}[h]
\begin{center}
\mbox{\includegraphics[width=10 cm,height=7 cm]{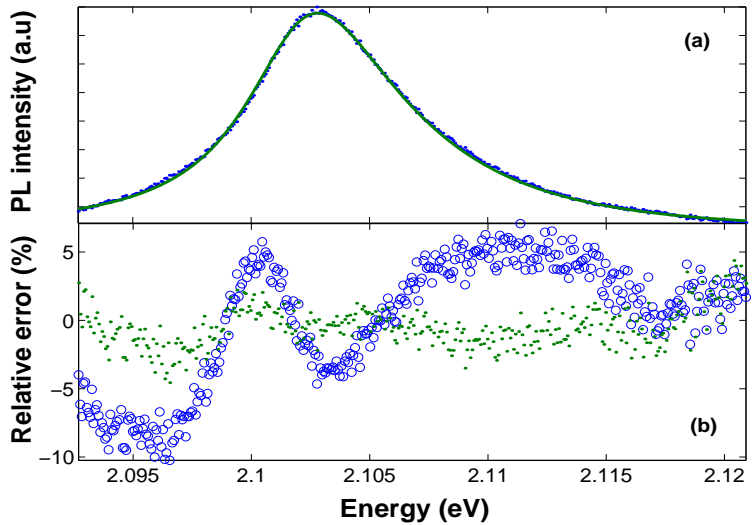}}
\caption{(a) Data and  fit using a 1D DOS of the D peak centered
at 2.102 eV for T=43 K. (b) Comparison of the error (relative to
the maximum signal value) between the fit function and the data
for the 1D model (points) and for the 2D model (open circles).
(From \cite{Dubin2002})} \label{raievib}
\end{center}
\end{figure}

The D emission line at T=43 K is shown in Fig.\ref{raievib}a. The
line-shape is clearly non lorentzian, this emission can then not
be that of a single state. In such a vibronic emission, the final
state involves a 1D optical phonon of the chain so that any
\textbf{k} state of the exciton band is connected to the ground
state by an optically allowed transition with generation of an
optical phonon with the same \textbf{k}. In Fig.\ref{raievib}a, a
fit barely distinguishable from the experimental data is presented
as well, and the corresponding error is given in
Fig.\ref{raievib}b. This fit has been obtained considering that
the energy dispersion of the optical phonons generated in the
vibronic emissions is very small compared to the one of the
excitons and is neglected. Since the effective mass of the exciton
is found to be $\approx$ 0.3 $m_{0}$ \cite{Lecuiller2002} ($m_{0}$
is the bare mass of the electron), this hypothesis seems very
reasonable. Moreover the transition matrix elements between all
the initial \textbf{k} exciton states and  the final state
(emission of one photon and one chain optical phonon) are assumed
equal (including the \textbf{k}=0 state): the contribution of the
initial \textbf{k} state to the overall homogeneous width  is the
same for all \textbf{k}.

The fitting function $f(E)$ is the convolution of the homogeneous
lorentzian profile and the population of emitting states (Eq.(
\ref{fitfunc})). This population is the product of the DOS by the
occupation probability. In (\ref{fitfunc}) $E_{0}$ is the
lorentzian's center position and $\Gamma_{vib}$ its half width.
$\it{A}$ is a constant including the amplitude of the lorentzian
and the constant parameters of the DOS. The only relevant
parameter in the fitting routine is $\Gamma_{vib}$.

\begin{equation}\label{fitfunc}
  f(E)=\frac{A}{(E-E_{0})^{2}+\Gamma_{vib}^{2}}\star
  \left(\exp(-\frac{E-E_{0}}{k_{B}T}).DOS \right)
\end{equation}

A quantitative 1D fit, i.e with a $(E-E_{0})^{-1/2}$ DOS, is
obtained for all vibronic lines at all temperature studied (5-50
K). Fitting with a 2D DOS is always worse. The 2D DOS corresponds
to the lowest dimensionality non singular density of states. The
fitting error for the 1D and 2D model are presented in
Fig.\ref{raievib}b, and show the very good accuracy of our model
when a 1D DOS is used.\

The analysis of the zero phonon emission line-width and its
temperature dependence ($\Gamma_{0}$(T)) allow the study of the
exciton-phonon interaction process. At low exciton densities and
low temperature the exciton-phonon scattering process is dominated
by acoustic phonons in semiconductors \cite{Basu}. Then, to
account for the thermal broadening of $\Gamma_{0}$, we have
studied the interaction between the 1D excitons confined on the
polymer chain and acoustic phonons. Previous calculations made by
Oh and Singh \cite{Singh2000} for quantum wells have been
adaptated to the 1D polymer chain. According to \cite{Singh2000},
since 3BCMU is a centro-symmetric crystal, the interaction with LA
phonons is the only one to be considered. Furthermore, 1D exciton
scattering by 1D LA phonon can just connect \textbf{k}=0 to
$k_{1D}$ ($k_{1D}$ is the solution of Eq.(\ref{1DLA})). This is
not in agreement with the experimental observation that all
\textbf{k} states emit within an energy range $k_{B}$T. Thus, we
have considered exciton interactions with the 3D LA phonons of the
monomer crystal.

This interaction derives from the deformation potential, and in
second quantization the resulting Hamiltonian in one dimension is
given by Eq.(\ref{H}). $K_{x}$ is the exciton momentum along the
chain axis and y,z denote the confined directions of chain.

\begin{equation}\label{H}
H_{ex-ph}(\overrightarrow{q}) = \sum_{K_{x}} C_{D}
[F^{-}(\overrightarrow{q})B^{+}_{K_{x}+q_{x}}B_{K_{x}}b_{q_{x}}+F^{+}(\overrightarrow{q})B^{+}_{K_{x}-q_{x}}B_{K_{x}}b^{+}_{q_{x}}]
\end{equation}
\begin{equation}
F^{-}(\overrightarrow{q}) = i
[D_{c}(q)u_{e}(q_{z})u_{e}(q_{y})G(\alpha_{h}q_{x})-D_{v}(q)u_{h}(q_{z})u_{h}(q_{y})G(-\alpha_{e}q_{x})]
\label{F}
\end{equation}

In (\ref{H}) $B^{+}_{K_{x}}$ ($B_{K_{x}}$) are the creation
(annihilation) operators of an exciton confined on the polymer
chain with a wave vector $K_{x}$. $b^{+}_{q_{x}}$ ($b_{q_{x}}$)
are the creation (annihilation) operators of an LA phonon with
momentum $q_{x}$ along the chain. $C_{D}=\sqrt{\frac{\hbar
q}{2\rho v_{s}V}}$ where V is the crystal volume, $\rho$ the
material density, q the modulus of the phonon momentum and $v_{s}$
the sound velocity. $\overrightarrow{q}=(q_{x},q_{y},q_{z})$ is
the phonon wave vector, and the subscripts + and - denote phonon
absorption and emission processes respectively. The exciton form
factor $F^{-}$ (note that $F^{- ~ *}$=$F^{+}$) is given by Eq.
(\ref{F}) where $\alpha_{h,e}=\frac{m_{h,e}}{m_{X}}$ (with
$m_{e,h,X}$ the electron, hole, exciton effective mass). $u_{e}$
($u_{h}$) are the form factors of the electron (hole) of the
exciton along the confined directions of the chain (Eq.
(\ref{u})), and $G_{e}$ ($G_{h}$) their form factors along the
chain axis (Eq.(\ref{G})).

\begin{eqnarray}
u_{i}(q_{j})= \int{ dj_{i}|\phi_{i}(j_{i})|^{2}e^{i q_{j}j_{i}}}
;i=e,h : j=y,z \label{u}
\\
G(\alpha q_{x})= \int{ dx e^{iq_{x}x}|\phi_{X}(x)|^{2}} \label{G}
\end{eqnarray}

$\phi_{e}$ and $\phi_{h}$ are the electron and hole wave function
for the lowest bound state for the charge carrier motion along the
confined axis of the polymer chain. $\phi_{X}$ is the exciton wave
function relative to its center of mass for the 1s state along the
chain axis x. $D_{c}(q)$ and $D_{v}(q)$ are the deformation
potential of the conduction and valence band respectively.

Finally, to quantify the thermal broadening of $\Gamma_{0}$, the
expression of the rate of a transition from $\textbf{k}$=0
involving one LA phonon is obtained by applying the Fermi Golden
rule. We note $|i>$ and $|f>$ the excitonic initial and final
states. The transition matrix element is given in Eq.(\ref{M}).

\begin{equation} \label{M}
M(K_{x} ->K_{x}+q_{x})=
<f|H_{ex-ph}(\overrightarrow{q})|i>=C_{D}^{2} |F^{+}|^2
f^{X}_{K_{x}}(f^{X}_{K_{x}+q_{x}}+1)n_{q}
\end{equation}

$n_{q}$ is the occupation number of a phonon with momentum q, and
$f^{X}_{K}$ is the occupation number of an exciton with momentum
K. The total rate for the transition from $K_{x}$ is given by:

\begin{equation}\label{Fermi}
W (K_{x})= \frac{2\pi}{\hbar} \sum_{q_{x}} |M(K_{x}
->K_{x}+q_{x})|^2 \delta(E(K_{x}+q_{x})-E(K_{x})-\hbar v_{s}q))
\end{equation}

The $\delta$ function in Eq.(\ref{Fermi}) induces a threshold in
the scattering due to the 1D character of the excitons. The
existence of this threshold does not depend on the dimensionality
of the phonons involved in the scattering. This scattering process
can be attributed to 1D LA phonons confined on the chain or 3D LA
phonons of the monomer. The argument of the delta function is
given by Eq.(\ref{3DLA}) for 3D phonons and by Eq.(\ref{1DLA}) for
1D phonons.

\begin{eqnarray}
E(K_{x}+q_{x})-E(K_{x})-\hbar
v_{s}q=\frac{\hbar^{2}q_{x}^{2}}{2m_{X}^{*}}-\hbar
v_{s,3D}\sqrt{q_{x}^{2}+q_{y}^{2}+q_{z}^{2}}\label{3DLA}
\\
E(K_{x}+q_{x})-E(K_{x})-\hbar
v_{s}q=\frac{\hbar^{2}q_{x}^{2}}{2m_{X}^{*}}-\hbar
v_{s,1D}q_{x}\label{1DLA}
\end{eqnarray}

\begin{figure}
\begin{center}
\mbox{\includegraphics[width=9 cm,height=6cm]{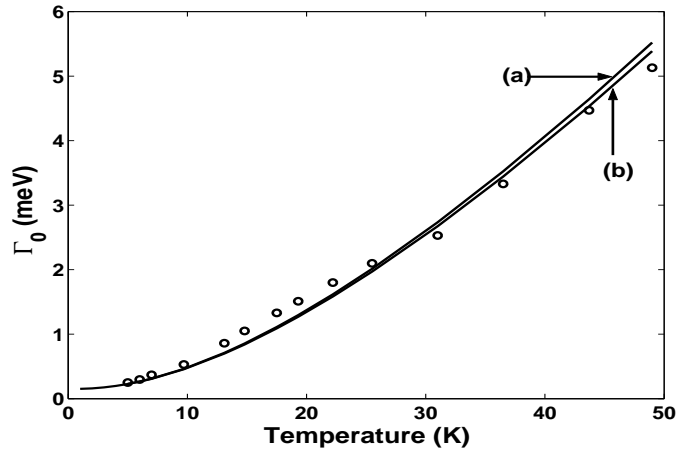}}
\caption{Experimental data (open circles) and calculated variation
of $\Gamma_{0}$ with temperature, for an exciton Bohr radius of 10
and 20 \AA ((a) and (b) respectively) and an effective mass of 0.3
$m_{0}$. In both cases, the sum of the deformation potential for
the valence and conduction band is $\approx$ 6 eV.}
\label{fitgamma0}
\end{center}
\end{figure}

$v_{s,1D}$  is the sound velocity along the chain and $v_{s,3D}$
the one in the DA 3BCMU monomer matrix assumed isotropic. In the
calculation the values measured in another (poly-)diacetylene,
(poly-)pTS \cite{vs,vs1} were used: $v_{s,1D}$= 5.5~$10^{3}\
m.s^{-1}$ and $v_{s,3D}$= 2.5~$10^{3}\ m.s^{-1}$ (the latter is
typical for molecular crystals). The difference between them is
due to the fact that the polymer is a chain of covalent bonds. As
mentioned above, from Eq.(\ref{1DLA}) one remarks that exciton
scattering by 1D LA phonon  can just connect \textbf{k}=0 to
$k_{1D}$. On the contrary scattering by 3D LA phonons of the
monomer matrix can connect \textbf{k}=0 to a continuum of \textbf
states with \textbf{k} $> \textbf{k}_{3D}^{min}$.  With
$m_{X}^{*}$ = 0.3 $m_{0}$ \cite{Lecuiller2002} one finds
$\textbf{k}_{3D}^{min}$= 8.2 $10^{-4} \AA^{-1}$. We also want to
note that in Eq.(\ref{3DLA}) the 3D LA phonon dispersion curve is
considered linear which is a reasonable approximation since the
\textbf{q} phonon states which contribute significantly to the
scattering are within the first tenth of the first Brillouin zone.
The variation of the transition rate is not highly dependent on
the exciton Bohr radius which is between 10 and 20 $\AA$
\cite{Sushai,Horvath} (see Fig. \ref{fitgamma0}). An exciton
effective mass $m^{*}_{X}=(0.3\ \pm 0.1)\ m_{0}$
\cite{Lecuiller2002} leads to typical deformation potential values
of $(D_{c}+D_{v})=6.1\ \mp 0.8$\ eV. The calculated scattering
rate goes to zero at 0 K so a constant parameter has been added in
order to reproduce experimental data. Its fitted value of
150~$\mu$eV is much larger than the contribution of the effective
lifetime of the exciton at 1 K (approximately 6 $\mu$eV)
\cite{Lecuiller2002}. This cannot be only due to instrumental
resolution and will be the subject of further analysis.

Thus, as shown in Figure \ref{fitgamma0} exciton scattering by 3D
LA phonons of the monomer matrix quantitatively explains the
thermal broadening of $\Gamma_{0}$. This thermalization process
has a characteristic time of 2 ps or less (see the widths in Fig.
\ref{fitgamma0}). Since the excitons have effective lifetimes over
100 ps in the range of temperatures studied \cite{Lecuiller2002},
excitons of the chain are in thermodynamic equilibrium with the
surrounding medium.

To summarize, we have presented micro-fluorescence experiments
performed on a single conjugated polymer chain in a crystalline
matrix. The zero phonon emission line is lorentzian while the
vibronic ones are asymmetric. Fitting the line-shape of these
vibronic peaks shows that the chain is a one dimensional system
which has to be described by an excitonic band with a $1/\sqrt{E}$
DOS. Furthermore,  the variation of $\Gamma_{0}$ with temperature
is explained by interactions with longitudinal acoustic phonons of
the 3D surrounding crystal in the range of temperatures studied.
This interaction thermalizes the excitons in their band so that
they are at thermodynamic equilibrium with the surronding medium.

The authors are thankful to P. Lavallard and G. Weiser for very
helpful discussions and to J. Kovensky for synthesizing the
diacetylene monomer. This research is supported by CNRS within the
program "nano-objet individuel", and by a grant from University
Denis Diderot.

\noindent

[*] Corresponding author: dubin@gps.jussieu.fr
\\


\begin{thebibliography}{99}
\bibitem{formule3B} The general formula of PDA is
$( =\! CR \! - \! C \! \equiv \! C \! - \! CR' \!=)_{n}$. R and R'
can be a large variety of chemical groups. In 3BCMU, R and R' are
identical with the following molecular formula:
-$(CH_{2})_{3}OCONHCH_{2}COOC_{4}H_{9}$
\end{thebibliography}
\end{document}